\documentclass[preprint,superscriptaddress]{revtex4-2}
\usepackage{chemformula}
\usepackage[T1]{fontenc}

\begin{document}

\newcommand{\dr}[1]{\textcolor{black}{#1}}
\newcommand{\ts}[1]{\textcolor{black}{#1}}
\newcommand{\ja}[1]{\textcolor{black}{#1}}
\newcommand{\todo}[1]{\textbf{\textcolor{black}{#1}}}

\author{D.~Rosenbach}\email[]{rosenbach@ph2.uni-koeln.de\\ present address: Physics Institute II, University of Cologne, 50937 Köln, Germany.}
\affiliation{Peter Gr\"unberg Institute (PGI-9), Forschungszentrum J\"ulich, 52425 J\"ulich, Germany}
\affiliation{JARA-Fundamentals of Future Information Technology, J\"ulich-Aachen Research Alliance, Forschungszentrum J\"ulich and RWTH Aachen University, Germany}

\author{A. R.~Jalil}
\affiliation{Peter Gr\"unberg Institute (PGI-10), Forschungszentrum J\"ulich, 52425 J\"ulich, Germany}
\affiliation{JARA-FIT Institute Green IT, RWTH Aachen University, 52062 Aachen, Germany}

\author{T. W.~Schmitt}
\affiliation{Peter Gr\"unberg Institute (PGI-9), Forschungszentrum J\"ulich, 52425 J\"ulich, Germany}
\affiliation{JARA-FIT Institute Green IT, RWTH Aachen University, 52062 Aachen, Germany}

\author{B.~Bennemann}
\affiliation{Peter Gr\"unberg Institute (PGI-10), Forschungszentrum J\"ulich, 52425 J\"ulich, Germany}
\affiliation{JARA-Fundamentals of Future Information Technology, J\"ulich-Aachen Research Alliance, Forschungszentrum J\"ulich and RWTH Aachen University, Germany}

\author{G.~Mussler}
\affiliation{Peter Gr\"unberg Institute (PGI-9), Forschungszentrum J\"ulich, 52425 J\"ulich, Germany}
\affiliation{JARA-Fundamentals of Future Information Technology, J\"ulich-Aachen Research Alliance, Forschungszentrum J\"ulich and RWTH Aachen University, Germany}

\author{P.~Sch\"uffelgen}
\affiliation{Peter Gr\"unberg Institute (PGI-9), Forschungszentrum J\"ulich, 52425 J\"ulich, Germany}
\affiliation{JARA-Fundamentals of Future Information Technology, J\"ulich-Aachen Research Alliance, Forschungszentrum J\"ulich and RWTH Aachen University, Germany}

\author{D.~Gr\"utzmacher}
\affiliation{Peter Gr\"unberg Institute (PGI-9), Forschungszentrum J\"ulich, 52425 J\"ulich, Germany}
\affiliation{JARA-Fundamentals of Future Information Technology, J\"ulich-Aachen Research Alliance, Forschungszentrum J\"ulich and RWTH Aachen University, Germany}



\author{Th.~Sch\"apers}
\affiliation{Peter Gr\"unberg Institute (PGI-9), Forschungszentrum J\"ulich, 52425 J\"ulich, Germany}
\affiliation{JARA-Fundamentals of Future Information Technology, J\"ulich-Aachen Research Alliance, Forschungszentrum J\"ulich and RWTH Aachen University, Germany}

\title{Ballistic surface channels in fully in situ defined Bi$_4$Te$_3$ Josephson junctions with aluminum contacts}

\date{\today}

\begin{abstract}
In this letter we report on the electrical transport properties of Bi$_4$Te$_3$ in a Josephson junction geometry using superconducting Al electrodes with a Ti interdiffusion barrier. Bi$_4$Te$_3$ is proposed to be a dual topological insulator, for which due to time-reversal and mirror symmetry both a strong topological insulator phase as well as a crystalline topological phase co-exist. The formation of a supercurrent through the Bi$_4$Te$_3$ layer is explained by a two-step process. First, due to the close proximity of the Al/Ti electrodes a superconducting gap is induced within the Bi$_4$Te$_3$ layer right below the electrodes. The size of this gap is determined by analysing multiple Andreev reflections (MARs) identified within the devices differential resistance at low voltage biases. Second, based on the Andreev reflection and reverse Andreev reflection processes a supercurrent establishes in the weak link region in between these two proximity coupled regions. Analyses of the temperature dependency of both the critical current as well as MARs indicate mostly ballistic supercurrent contributions in between the proximitized Bi$_4$Te$_3$ regions even though the material is characterized by a semi-metallic bulk phase. The presence of these ballistic modes gives indications on the topological nature of Bi$_4$Te$_3$.
\end{abstract}


\maketitle

\section{Introduction}

Hybrid structures of three-dimensional topological insulators and superconductors are considered promising building blocks for the realization of topological quantum circuits \cite{Hassler2011,Cook2012,Hyart2013}. A crucial optimization parameter is a sufficiently large coupling of the superconductor to the topological insulator. In order to probe the proximity coupling strength a Josephson junction with a topological insulator weak link bridging two superconducting electrodes can be employed \cite{Dominguez2012, Wiedenmann2015, Bocquillon2016}. By measuring the current-voltage characteristics of these junctions, the interface transparency as well as the underlying mode of transport, i.e. diffusive or ballistic, can be investigated \cite{Schueffelgen2019, Rosenbach2021}. The supercurrent in a Josephson junction is carried by electron-hole bound states. Based on the nature of these bound states their energy phase relation (E$\Phi$R) has a fixed periodicity. Irradiating the junction with a radio frequency signal allows to investigate the Shapiro step response. As the Josephson voltage $V_0 = hf/2e$ in between two Shapiro steps depends on the periodicity of the the bound states E$\Phi$R they give indications on the nature of the bound states \cite{Dominguez2012}. In junctions with topological weak link both Andreev bound states (ABS; diffusive bulk and surface modes) carrying $2e$ charge per cycle and Majorana bound states (MBS; ballistic, perfectly transmitted surface modes) carrying only a single $1e$ charge per cycle coexist. Hence the periodicity of the bound states E$\Phi$R and respective the Josephson voltage in between two Shapiro steps differ by a factor of $2$ comparing MBSs to ABSs \cite{Dominguez2017}. MBSs are crucial to probe the existence of zero energy states within topological Josephson junctions and are indicated by missing odd Shapiro steps in experiments \cite{Wiedenmann2015,Bocquillon2016,Schueffelgen2019}.\\
  
Topological Josephson junctions can be separated into two regions. The first region is the topological matter underneath the superconducting electrodes. Here, the proximity effect opens an effective induced superconducting gap within both the surface and bulk of the topological matter. The second region is in between these two laterally separated proximitized regions called the weak link defined by the non-proximitized part of the topological matter. For the investigation of novel topological matter the question arises what relevant transport channels exist and what is their main mode of transport, i.e. ballistic topological surface states or diffusive bulk states.\\
For the weak link in between the superconducting electrodes we chose Bi$_4$Te$_3$, which is a natural superlattice of alternating Bi$_2$ bilayers \cite{Akturk2016} and Bi$_2$Te$_3$ quintuple layers. Initially, Bi$_4$Te$_3$ has been reported to be a zero band gap semimetal, comprising a Dirac cone at the $\Gamma$-point \cite{Saito2017}. More recently, band structure calculations supplemented with scanning tunneling spectroscopy and angular photoemission spectroscopy measurements showed that Bi$_4$Te$_3$ is a semimetal with topological surface states \cite{Chagas2020,Chagas2022,Nabok2022}. In advanced $GW$-band structure calculations a band gap of about $0.2$\,eV was identified around the $\Gamma$-point. Owing to time-reversal and mirror symmetries, Bi$_4$Te$_3$ is a strong topological insulator (STI) as well as a topological crystalline insulators (TCI). Furthermore, it is predicted that it contains higher order topological states \cite{Nabok2022,Jalil2022}.\\  

We report on the transport properties of Josephson junctions based on the Bi$_4$Te$_3$ material system as weak link material and Al/Ti as the superconducting electrodes. For the fabrication of the samples, we employed an all in situ method \cite{Schueffelgen2019, Rosenbach2020,Schmitt2022}, meaning that the Bi$_4$Te$_3$ weak link layer is grown by selective-area molecular beam epitaxy, while for the definition of the superconducting Al/Ti electrodes we use an in situ shadow evaporation technique. This approach allows to achieve a clean interface between the Bi$_4$Te$_3$ layer and the superconductor without any contamination. In our study the proximity strength of the Al/Ti superconducting electrodes towards the underlying Bi$_4$Te$_3$ nanoribbon is examined in low temperature transport experiments including current-voltage characteristics and differential resistance measurements. From multiple Andreev reflections (MARs) identified within the differential resistance we gain information about the strength of the proximity effect in Bi$_4$Te$_3$ and the size of the induced superconducting gap. Furthermore, from the excess current and from the temperature dependence of both the junctions critical current and the MARs we are able to specify the dominant transport regime of the Josephson supercurrent. 

\section{Experimental Setup}
Nanoribbon Josephson junctions have been defined following an all in situ approach\cite{Schueffelgen2019, Rosenbach2020,Schmitt2022}. Therefore, two independent masking techniques are used. The masks are defined using four alternating layers of SiO$_2$ and Si$_3$N$_4$ deposited on a highly resistive Si (111) substrate ($R > 2000\,\Omega\cdot $cm) \cite{Schmitt2022}. The first two layers are 5\,nm of oxidized SiO$_2$ and 15\,nm of low pressure chemical vapor deposited (LPCVD) Si$_3$N$_4$. They comprise the selective area growth (SAG) mask. Narrow ($w=1000\,$nm down to $100\,$nm) nanotrenches are etched into the top Si$_3$N$_4$ layer using a combination of electron beam lithography and reactive ion etching. Afterwards, a 300-nm-thick SiO$_2$ layer and a 100-nm-thick Si$_3$N$_4$ layer are deposited using LPCVD. These layers comprise the stencil mask used to deposit the superconducting electrodes in situ. Free-hanging Si$_3$N$_4$ bridge structures are defined, as previously reported \cite{Schueffelgen2019}, by patterning the Si$_3$N$_4$ and subsequently removing the SiO$_2$ underneath in hydrofluoric acid (HF). The HF dip also locally removes the SiO$_2$ of the first oxidized layer of the SAG mask only within the Si$_3$N$_4$ nanotrenches. During molecular beam epitaxy the Bi$_4$Te$_3$ will selectively grow within these nanotrenches on top of the Si(111) that is revealed during SiO$_2$ removal. The free-hanging Si$_3$N$_4$ bridge structures will be used after the deposition of Bi$_4$Te$_3$ to define the superconducting electrodes, without breaking the vacuum \cite{Schueffelgen2019}.\\
\begin{figure*}[!b]
	\centering
\includegraphics[width=\textwidth]{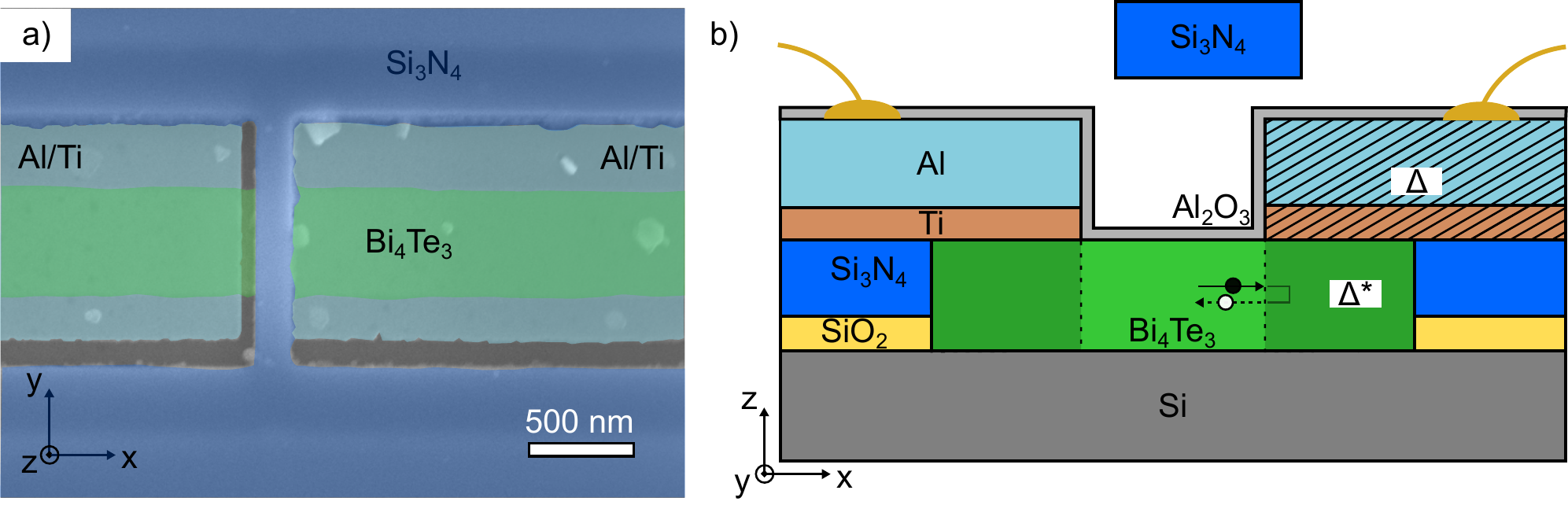}
	\caption{\textbf{In situ deposited Bi$_4$Te$_3$ nanoribbon Josephson junction}. a) shows a false-colored SEM graph of the top view of an Al/Ti - Bi$_4$Te$_3$ - Ti/Al Josephson junction as prepared in situ. The Al/Ti superconducting electrodes are highlighted in cyan/brown, while the Bi$_4$Te$_3$ nanoribbon is shown in green and the Si$_3$N$_4$ hard mask in blue. The cross section along the nanoribbon main axis is schematically depicted in b). Here, the Ti interdiffusion layer (brown), the Al$_2$O$_3$ dielectric capping layer (light grey), the Si substrate (dark grey) as well as the Si$_3$N$_4$/SiO$_2$ (blue/yellow) SAG mask layers are visible. The Al/Ti contacts are attributed a composite superconducting pair parameter $\Delta$ and the pair parameter of the proximity coupled region in the Bi$_4$Te$_3$ layer (dark green) is denoted by $\Delta^{*}$.}
	\label{fig1}
\end{figure*}
 
Bi$_4$Te$_3$ is a stoichiometric state of the (Bi$_2$Te$_3$)$_m$(Bi$_2$)$_n$ family with $(m:n) = (3:3)$. A unit cell comprises an alternating stacking sequence of Bi$_2$Te$_3$ quintuple layers and Bi bilayers. The planar epitaxy of Bi$_4$Te$_3$ stoichiometric alloy is achieved via molecular beam epitaxy (MBE) by precisely controlling the Bi:Te beam flux ratio to 1:2 while keeping $T_\mathrm{Bi}$ at $490^\circ$C and $T_\mathrm{Te}$ at $280^\circ$C \cite{Jalil2022}. In order to acquire Bi$_4$Te$_3$ nanostructures, the optimum growth parameters are subjected to the pre-patterned substrates with combinational surfaces. The substrate rotation ensures a homogeneous growth of the Bi$_4$Te$_3$ layer also underneath the free-hanging Si$_3$N$_4$ bridges. The thickness of the Bi$_4$Te$_3$ nanoribbon depends on the geometry and width of the nanotrenches \cite{Jalil2022}. This is as also adatoms impinging on the Si$_3$N$_4$ within the limit of the adatom diffusion length can contribute to the growth of Bi$_4$Te$_3$ within the trenches. For the junctions investigated here their respective thicknesses are given in Tab.~\ref{Tab1}. The superconducting electrodes are deposited within a nitrogen cooled chamber below 0$^\circ$C by turning off the substrate heater. The free-hanging Si$_3$N$_4$ bridges are aligned perpendicular to the effusion cells of the evaporated metal, such that the shadow defines the weak link area. After electrode deposition devices are covered by a 5\,nm thin Al$_2$O$_3$ dielectric layer electron beam evaporated from a stoichiometric target. A false-colored scanning electron micrograph of an as-prepared Josephson junction with aluminum superconducting contacts is shown in Fig.~\ref{fig1} a). Aluminum has previously been reported to diffuse into (Bi$_{0.06}$Sb$_{0.94}$)$_2$Te$_3$ thin films \cite{Schueffelgen2017}, which increases the interfacial resistance of junctions within the superconducting regime of the Al electrodes. In order to prevent diffusion of the Al into the underlying Bi$_4$Te$_3$ layer, a 3\,nm thin Ti layer is deposited first as an interdiffusion barrier, as depicted in the schematics of the junction cross section shown in Fig.~\ref{fig1} b). The critical temperature of the superconducting Al/Ti composite electrodes is determined to be $T_{\text{c}}=0.95\,$K from four-terminal measurements of the differential resistance as a function of the temperature $T$ down to 23\,mK base temperature of a dilution refrigerator. The magnitude of the superconducting pair parameter has been determined to measure $\Delta=145\,\mu$eV, following Bardeen-Cooper-Schrieffer theory \cite{Bardeen1957}.\\
\begin{table}[!t]
\begin{center}
\begin{tabular}{| c | c | c | c | c | c | c | c | c | c | c | c | c | }
\hline
\# & $w$ [nm] & $L$ [nm]& $\overline{t}$ [nm] & $I_c$ [nA] & $R_{N}$ [$\Omega$] & $I_cR_N$ [$\mu$eV] & $\Delta^*$ [$\mu$eV] & $I_{\text{exc}}$ [nA] & $I_{\text{exc}}R_N$ [$\mu$eV] & $\alpha$ & $\tau$ & $\gamma_B$\\
\hline
\hline
1 & 1000 & 130 & 8.6 & 176 & 120 & 21.12 & 82.5 & 500 & 60 & 0.72 & 0.65 & 0.52 \\
\hline
2 & 500 & 130 & 10 & 35 & 310 & 10.85 & 95 & 159 & 49.3 & 0.5 & 0.57 & 0.36 \\
\hline
3 & 100 & 140 & 16.5 & 30 & 744 & 22.32 & 115 & 75 & 55.8 & 0.49 &  0.56 & 0.24 \\
\hline
\end{tabular}
\caption{Overview of interface parameters of Josephson junctions with Bi$_4$Te$_3$ nanoribbon weak link and Al/Ti (30\,nm/3\,nm) superconducting contacts. Given are the geometric dimensions, the junction width $w$, the junction length/electrode separation length $L$ and the mean thickness $\overline{t}$ of the nanoribbon. The proximity induced gap below the superconducting electrodes $\Delta^*$, the excess current $I_c$ as well as the dimensionless parameters $\alpha$, $\tau$ and $\gamma_B$ that describe the interfacial quality of the junctions.}
\label{Tab1}
\end{center}
\end{table}

The electrodes of the in situ defined nanoribbon Josephson junctions are wire bonded to a chip carrier in a quasi-four terminal contact configuration. The junctions are cooled to a base temperature of $T=23\,$mK using a dilution refrigerator. At base temperature the sample resistance is measured using standard lock-in techniques and the potential drop across the junction is determined using a voltmeter. 

\section{Experimental Results}
\subsection{Multiple Andreev reflections}
\begin{figure*}[!ht]
	\centering
\includegraphics[width=\textwidth]{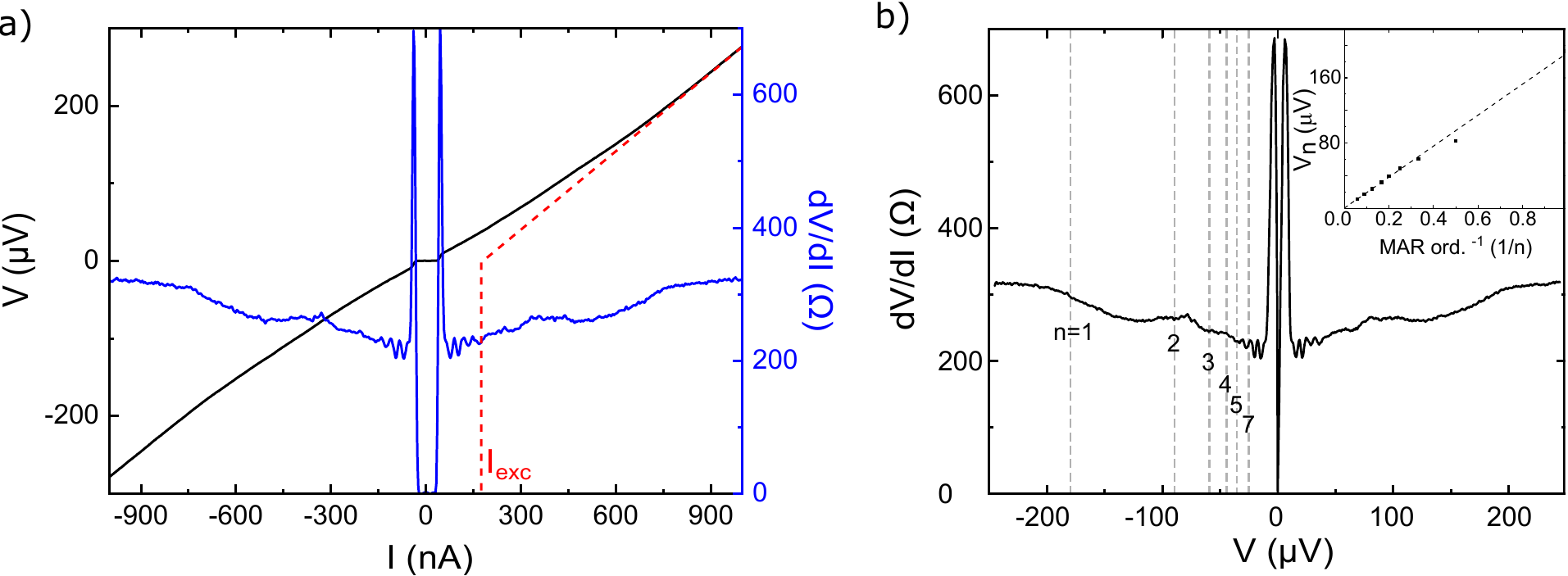}
	\caption{\textbf{$IV$-characteristics and differential resistance $dV/dI$ of Josephson junction \#2.} a) $IV$-characteristics and differential resistance as a function of the applied d.c. bias current ($dV/dI(I)$). A linear extrapolation from the IV-characteristics above $2\Delta^*$ to $V=0$ is shown (red dashed line) to extract the excess current $I_{\text{exc}}$. b) Differential resistance as a function of the measured d.c. potential drop across the Josephson junction ($dV/dI(V)$), showing signatures of multiple Andreev reflections (MARs). The inset shows the position ($V_n$) of the MARs plotted against the inverse of the MAR order number (1/n). The linear fit is forced through the origin.}
	\label{fig2}
\end{figure*}

We have measured three junctions of different width $w=100, 500$, and $1000\,$nm (see Tab.~\ref{Tab1}). Both $dV/dI (I)$ as well as $V (I)$ as a function of a d.c. current bias applied for the 500,nm wide junction (junction \#2) are shown in Fig.~\ref{fig2} a). For an applied bias current below the critical current of $I_c =35\,$nA (see also Tab.~\ref{Tab1}) a Josephson supercurrent establishes. The differential resistance $dV/dI$ is zero below $I<I_c$ and reaches a finite value as soon as the current bias exceeds $I>I_c$. The critical current is not hysteretic, whether the current bias is swept from positive to negative current biases or vice versa. In Tab.~\ref{Tab1} the critical current $I_c$, the normal state resistance $R_N$ values are listed. The corresponding $I_cR_N$ product values are found to be in the range between $10.85$ and $22.32\,\mu$V.\\ 
As mentioned before, we anticipate that establishing a supercurrent through the Bi$_4$Te$_3$ layer is a two step process. First, the proximity to the superconducting metallic Al/Ti electrodes induces a superconducting pair potential into the Bi$_4$Te$_3$ layer (dark green regions in Fig.~\ref{fig1} b)), which decays over a length scale given by the superconducting coherence length $\xi_{\text{N}}$ within the Bi$_4$Te$_3$ layer. For the superconducting coherence length we have to consider two different cases. In the 'dirty limit', the elastic scattering in the dissipative state of the Bi$_4$Te$_3$ layer takes place on length scales smaller than the superconducting coherence length. When the distance between two elastic scattering events exceeds the superconducting coherence length, the transition is in the 'clean limit'. Using low-temperature magnetotransport data on nano-Hall structures, we find that the Bi$_4$Te$_3$ layer is (semi)metallic, in agreement with recent reports \cite{Chagas2020,Chagas2022,Nabok2022}, with a carrier density of $n_{\text{2D}} \approx 4\times 10^{14}\,$cm$^{-2}$ (see Supplementary Sec.~A) and an elastic mean free path length of only $l_e \approx 4\,$nm. Furthermore, the Hall bar data does not show any significant increase of the magnetoresistance, as expected from a Dirac semimetal \cite{Liang2014}. For given reasons we therefore assume that the proximitized regions of the Bi$_4$Te$_3$ film underneath the superconducting Al/Ti electrodes are in the dirty limit, since the estimated superconducting coherence length of $\xi_{\text{N}}=\sqrt{\hbar D_{\text{Bulk}}/2\pi k_B T_c}=45\, $nm, with $D_{\text{Bulk}}$ the diffusion constant of the bulk and $T_c$ the critical temperature of the Ti/Al superconducting electrodes.\\

When proximitizing the regions of the Bi$_4$Te$_3$ underneath the Al/Ti superconducting electrodes a Josephson supercurrent establishes in a next step between the two proximitized layers based on electron-hole bound states. When the applied current bias exceeds the critical current $I>I_c$ the junctions resistance is modulated by Andreev reflection processes at the superconductor to normal conductor interface. Only beyond a current bias of about $|I | >740\,$nA the junctions resistance is mostly constant. At this point the potential drop across the junction measures $2\Delta^*$, i.e. the size of the proximity induced superconducting gap, as indicated in Fig.~\ref{fig1} b). In order to quantify the size of the induced superconducting gap $\Delta^*$ in the proximitized Bi$_4$Te$_3$ more precisely (cf. Fig.~\ref{fig1} b)) we analyzed multiple Andreev reflections (MARs) visible within the differential resistance $dV/dI$ of the junction. These MARs occur at bias voltages below the size of the induced superconducting gap at voltages of $V = 2\Delta^*/en$, where $n$ is an integer \cite{Kleinsasser1994}.  In junction \#2 we observe MARs of the order $n=1,2,3,4,5,6,9,11,13$. Missing signatures of intermediate order MARs (e.g. $n=7,8$) has been observed before in BiSbTeSe$_2$ nanoribbon Josephson junctions \cite{Jauregui2018} but an explanation is missing until now.\\
The size of the induced superconducting gap is determined by plotting the position (in volt) of each MAR against the inverse of the MAR order number (1/n). The induced superconducting gap measures $\Delta^*=95\,\mu$eV (for $n=1$, as indicated by a blue dot within Fig.~\ref{fig2} b) at $T=50\,$mK), which is smaller than the gap of the Al/Ti superconducting electrodes ($\Delta=145\,\mu$eV). \\

\begin{figure*}[!b]
	\centering
\includegraphics[keepaspectratio=true, width=.42\textwidth]{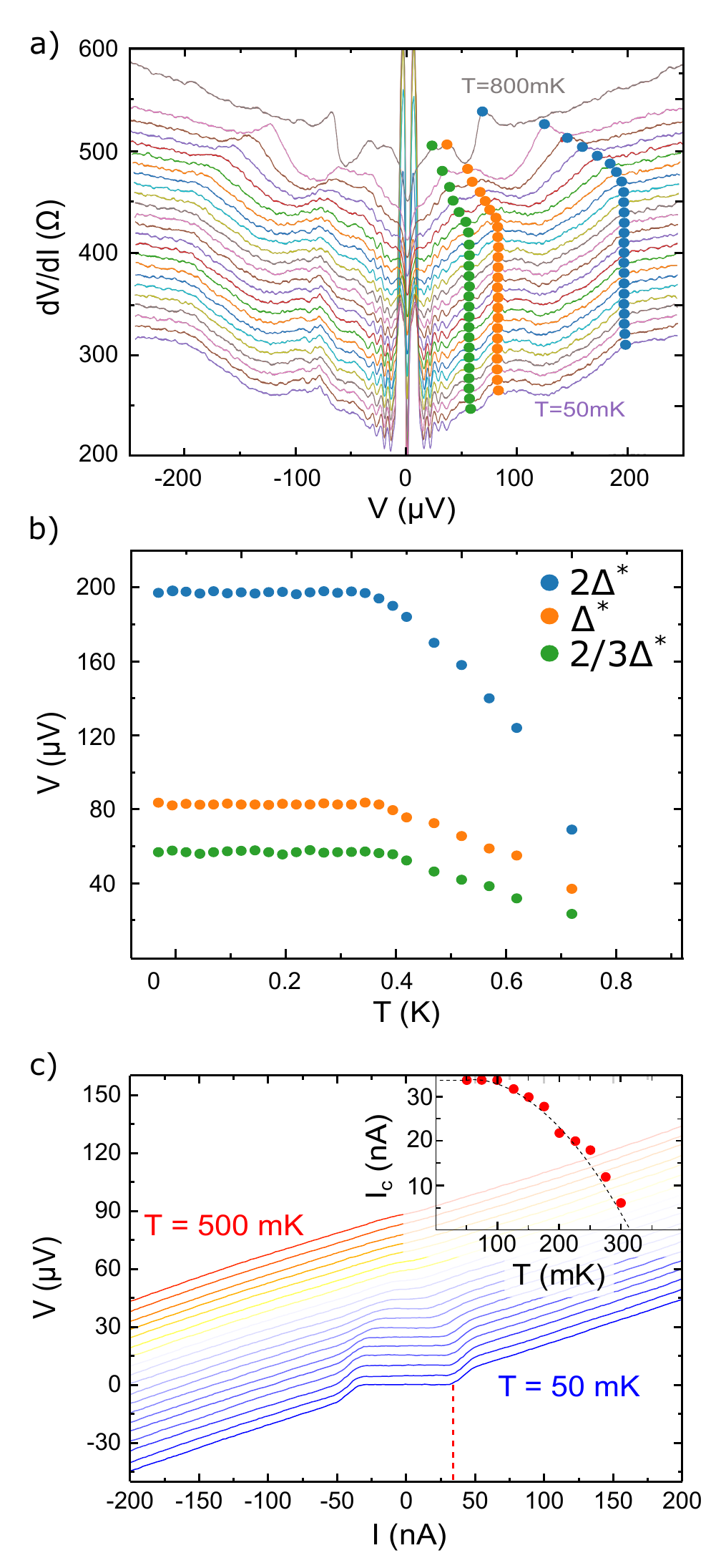}
	\caption{\textbf{Temperature dependency} of  a) the differential resistance as a function of the bias potential of Josephson junction \#2, $dV/dI (V,T)$, in between 50\,mK and 800\,mK. Each trace is offset by 25\,$\Omega$. The size of the proximity induced superconducting gap (2$\Delta^*$ in $\mu$V) is highlighted by blue dots, while MARs of order $n=2$ and $3$ are highlighted by orange and green dots respectively. b) shows the temperature dependent position of 2$\Delta^*$ and MARs of order $n=2$ and $n=3$. c) shows the temperature dependent $IV$-characteristics from 50\,mK to 500\,mK, where each trace is offset by $5\,\mu$V. The red dashed line indicates the critical current of the junction $I_{\text{c}}=35\,$nA at $T=50\,$mK. The temperature dependent critical current, $I_{\text{c}}(T)$, is shown in the inset with the fit indicated by the dashed line.}
	\label{fig3}
\end{figure*}

\subsection{Temperature dependency of $I_c$ and MARs}

Figure~\ref{fig2} b) shows the differential resistance of junction \#2 at different temperatures. The signatures of MARs vanish above the critical temperature of the Al/Ti superconducting electrodes. The temperature dependency of MARs of order $n=1$ (blue dots), $n=2$ (orange dots) and $n=3$ (green dots) is shown in Figs.~\ref{fig2} b) and c). The temperature dependency of the induced superconducting gap is given by \cite{Aminov1996}
\begin{equation}
    \Delta^* (T) = \frac{\Delta_{\text{Al/Ti}}(T)}{1+\gamma_{\text{B}}\sqrt{\Delta^2_{\text{Al/Ti}}(T)-\Delta^{*2}(T)}/k_{\text{B}}T_c} \, ,
\end{equation}
where $\gamma_B$ is a measure of the interfacial barrier strength in between the Al/Ti superconducting electrodes and the Bi$_4$Te$_3$ nanoribbon layer. Above formula is fitted to the $\Delta^* (T)$ data and a value for $\gamma_B=0.36$ has been determined. The $\gamma_B$ values of the other two junctions are listed in Tab.~\ref{Tab1}. The value of $\gamma_B$ indicates that there is an effective barrier present between the Al/Ti layer and Bi$_4$Te$_3$ despite the in situ fabrication. It has been identified that the Bi$_4$Te$_3$ tends to be terminated by a Bi bi-layer underneath the Al/Ti layer while it is terminated by a Bi$_2$Te$_3$ layer otherwise \cite{Jalil2022}. A possible reason for the barrier identified might be the mismatch of Fermi energies in between these different regions on the surface of the proximitized and the non-proximitized regions of Bi$_4$Te$_3$ resulting in a potential step at their interface. \\

As a next step, the Josephson supercurrent between the the proximitized regions with the superconducting gap $\Delta^*$ is analyzed in detail. The supercurrent depends on the kind of transport, i.e. ballistic or diffusive, and on the transparency between the proximitized Bi$_4$Te$_3$ layers and the Bi$_4$Te$_3$ weak link (cf. Fig.~\ref{fig1} b). The transparency of the interfaces of the lateral Josephson junction are analyzed in two ways. The first method uses the excess current of $I_{\text{exc}}=159\,$nA, which is determined from the junctions $IV$-characteristics by linear extrapolation above the superconducting gap $V\geq 2\Delta^*$ (highlighted as dashed red line in Fig.~\ref{fig2} a)) to $V=0$. The excess current displays the additional current due to successful Andreev reflections in the dissipative state of the junction and is directly related to the transparency of the junction. \ts{An analytical expression following Niebler, Cuniberti, and Novotny \cite{Niebler2009} is used to determine the junctions interfacial barrier strength $Z=0.86$ using the parameter $\alpha= eI_{\text{exc}} R_N/\Delta^*$ (cf. Tab.~\ref{Tab1}). The barrier strength is related to the transparency via $\tau=1/(1+Z^2)=0.57$.} Note, that $\tau$ expresses the transparency between the proximitized and the non-proximitized regions of the Bi$_4$Te$_3$ layer in contrast to $\gamma_B$ which quantifies the barrier between the Al/Ti superconducting electrodes and the proximitized Bi$_4$Te$_3$ region. Similar to the observation from Kunakova et al.\cite{Kunakova2019} we find that the values for the interface transparency parameters $\tau$ and $\gamma_B$ are interdependent. This effect can be attributed to a metallization effect the electrodes have on the surface states of the Bi$_4$Te$_3$ layer.\\
 
An additional method to determine the interface transparency is by quasi-classical analysis of the temperature dependent critical current \cite{Schueffelgen2019, Rosenbach2021}. Using a voltage criterion the critical current is extracted from the $IV$-characteristics at different temperatures shown in Fig.~\ref{fig2} d), with the inset showing $I_c(T)$. We used a ballistic model fit (shown as black dashed line in the inset of Fig.~\ref{fig1} b)) based on the Gor’kov equations with arbitrary junction length $L$ \cite{Brinkman2000} and barrier transparency $\mathcal{D}$ \cite{Galaktionov2002}. For the fit a value for the critical temperature of the gap within the Al/Ti electrodes $T_c = 0.95\,$K and a Fermi velocity of $v_F = 3.8\times10^5\,$m/s of the Bi$_2$Te$_3$ surface layer are used \cite{Jalil2022}. The best fit results in an interface transparency of $\mathcal{D}=0.6$, which is in good agreement with the transparency $\tau$ determined using the excess current analysis described before. We also performed a quasi-classical fit using a diffusive model based on the Usadel equations \cite{Usadel1970}. However, within a physically reasonable range of values we did not get a decent fit. Our analysis indicates that the supercurrent is carried by ballistic modes with increased superconducting coherence length rather than bulk modes. The superconducting coherence length of these ballistic modes can be estimated within the clean junction limit to measure $\xi_{\text{N}} = \hbar v_{\text{F}}/2\pi k_{\text{B}}T_c=1.15\,\mu$m. The observation of a dominating ballistic channel might be attributed to highly conductive surface states of the Bi$_4$Te$_3$ layer, which overrules the diffusive transport in the bulk.\\

\subsection{Shapiro steps}
\begin{figure*}[!ht]
	\centering
\includegraphics[width=0.8\textwidth]{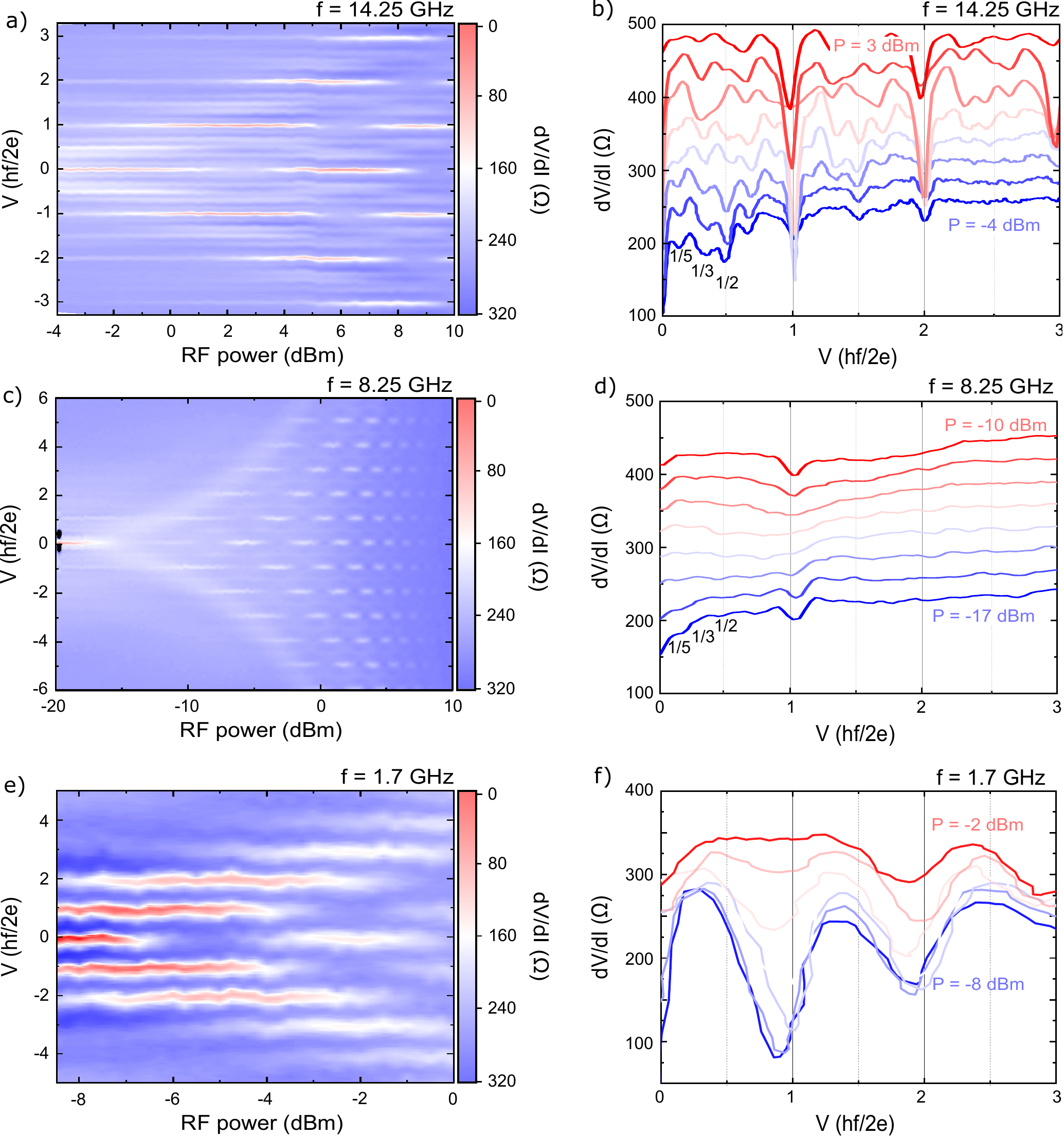}
	\caption{\textbf{Shapiro response of Josephson junction \#2 at different radio-frequencies applied.} a), c) and e) show the differential resistance as a function of the radio-frequency excitation amplitude/radio-frequency power ($P$ in dBm) and the potential bias ($V$ in $hf/2e$) of the junction ($dV/dI (P,V)$). The differential resistance is displayed in between values of $dV/dI=0$ (red) and $dV/dI=320\,\Omega$. b), d) and f) show line traces of the differential resistance as a function of the bias potential in a given range of radio-frequency powers, with the lowest power displayed in blue and the highest in red. Next to Shapiro steps at full integer values of $V=n\cdot hf/2e$, there are sub-integer steps visible in line-cuts at $f=14.25\,$GHz and $8.25\,$GHz. The sub-integer steps are highlighted with their given fractions of the first integer Shapiro step.}
	\label{fig4}
\end{figure*}

We also performed differential resistance measurements under the influence of an externally applied radio-frequency (rf) signal using a $\lambda/4$ antenna. In Figs.~\ref{fig4} a)-f) the differential resistance is displayed as a function of the applied rf power and the junctions potential difference is scaled by $hf/2e$. Within the range of frequencies applied $1.7\,$GHz $\leq f_{\text{rf}} \leq 14.25\,$GHz we observe full integer Shapiro steps. At frequencies $f_{\text{rf}}=14.25\,$GHz and $8.25\,$GHz, however, additional sub-integer Shapiro steps have been measured. Fractional Shapiro steps can be caused by phase-slip centers inside the junction \cite{Ivlev1984}, phase instabilities introduced by Abrikosov vortices \cite{Golubov2004}, magnetic disorder \cite{Sellier2004} or due to a non-sinusoidal or skewed current phase relation (C$\Phi$R) \cite{Panghotra2020,Raes2020}. In junctions of high transparencies or very short ballistic junctions the C$\Phi$R is expected to be non-sinusodial \cite{Golubov2004}. The relative measure of the junction length over the superconducting coherence length within the Bi$_4$Te$_3$ layer ($d/\xi_{\text{N}}$) has influence on the maximum Josephson supercurrent and the shape of the C$\Phi$R. Already for values of $\xi_{\text{N}}/d > 3$ the C$\Phi$R is skewed and the maximum current density lies above a value of $\phi$ > $\pi$ \cite{Golubov2004}. Skewed, non-sinusoidal C$\Phi$Rs can be decomposed into sinusoidal components of lower periodicity, which can explain the evolution of sub-integer Shapiro steps. For ballistic modes of increased superconducting coherence length ($\xi_{\text{N}} = 1.15\,\mu$m) this limit would need to be considered as the junction length of junction \#2 ($L=130\,$nm) is much smaller. For the wider junction (junction \#1, $w=1000\,$nm) as for the narrower junction (junction \#3, $w=100\,$nm, see supplementary Sec.~B) the sub-integer Shapiro steps have been observed as well, confirming the presence of ballistic modes independent of the junction geometry.\\
For the presence of ballistic modes one would expect the supercurrent to be partially carried by MBSs, resulting in odd integer Shapiro steps to vanish \cite{Dominguez2017}. Based on the fraction of the supercurrent that is carried by MBSs compared to the supercurrent carried by ABSs the cross-over frequency for the observation of missing odd integer Shapiro steps \cite{Dominguez2012} should lie below $f_{MBSs}<5.25\,$GHz, which is the cross over frequency considering the whole supercurrent is carried only by MBSs. As no missing Shapiro steps have been recorded it is assumed that less then one third of the supercurrent is carried by MBSs. Therefore, a possible reason that we did not observe missing odd integer Shapiro steps as an indication of MBSs might be that the irradiated frequency was too large \cite{Dominguez2017}.\\

\section{Conclusions}

By characterizing Bi$_4$Te$_3$-based Josephson junctions we obtained a detailed picture of the different contributions taking part in establishing a supercurrent through the junctions weak link. By analysing MARs we found that the intimate contact of the Al/Ti layer on top of the Bi$_4$Te$_3$ layer results in an induced superconductive gap $\Delta^*$ in the topological matter due to the proximity effect. In order to establish robust proximitzed regions underneath the Al/Ti electrodes the presence of bulk carriers are probably beneficial, if not essential. The Bi$_4$Te$_3$ has been identified to carry a large amount of bulk charges. The proximitized regions of the Bi$_4$Te$_3$ are coupled by the unproximitized Bi$_4$Te$_3$ weak link giving rise to a Josephson supercurrent. We anticipate that the Josephson supercurrent mainly flows in the topological surface channel rather than in the bulk of the Bi$_4$Te$_3$ link, similar to results obtained in junctions with a different topological insulator layer \cite{Schueffelgen2019}. Analysing the temperature dependency of the critical current we indeed identified the transport regime in these junctions to be mainly ballistic. However, by analysing the temperature dependency of the MARs an effective barrier in between these regions, probably due to a different surface termination of both regions, has been identified. From our Shapiro step measurements we came to the conclusion that the current-phase relationship is non-sinusoidal, i.e. supporting our claim of ballistic modes in our junctions. However, we did not find a 4$\pi$ contribution in the Shapiro step measurements indicating the presence of Majorana zero modes. One reason might be that our lowest rf frequency of $f \leq 1.7\,GHz$ was too high, so that we could not enter the regime where the 4$\pi$ contributions are visible. For future material combinations in hybrid Josephson junctions including topological matter it is important to consider our findings. 
 
\section*{Acknowledgements}

 This work was partly funded by the Deutsche Forschungsgemeinschaft (DFG, German Research Foundation) under Germany's Excellence Strategy - Cluster of Excellence Matter and Light for Quantum Computing (ML4Q) EXC 2004/1 - 390534769. This work was financially supported by the German Federal Ministry of Education and Research (BMBF) via the Quantum Futur project "MajoranaChips" (Grant No. 13N15264) within the funding program Photonic Research Germany.\\

\textbf{Competing interests}\\
The author(s) declare no competing interests.\\

\bibliography{Bibliography}

\newpage

\section*{Supplementary Information}

\subsection{Magnetotransport}
\begin{figure*}[!ht]
	\centering
\includegraphics[width=\textwidth]{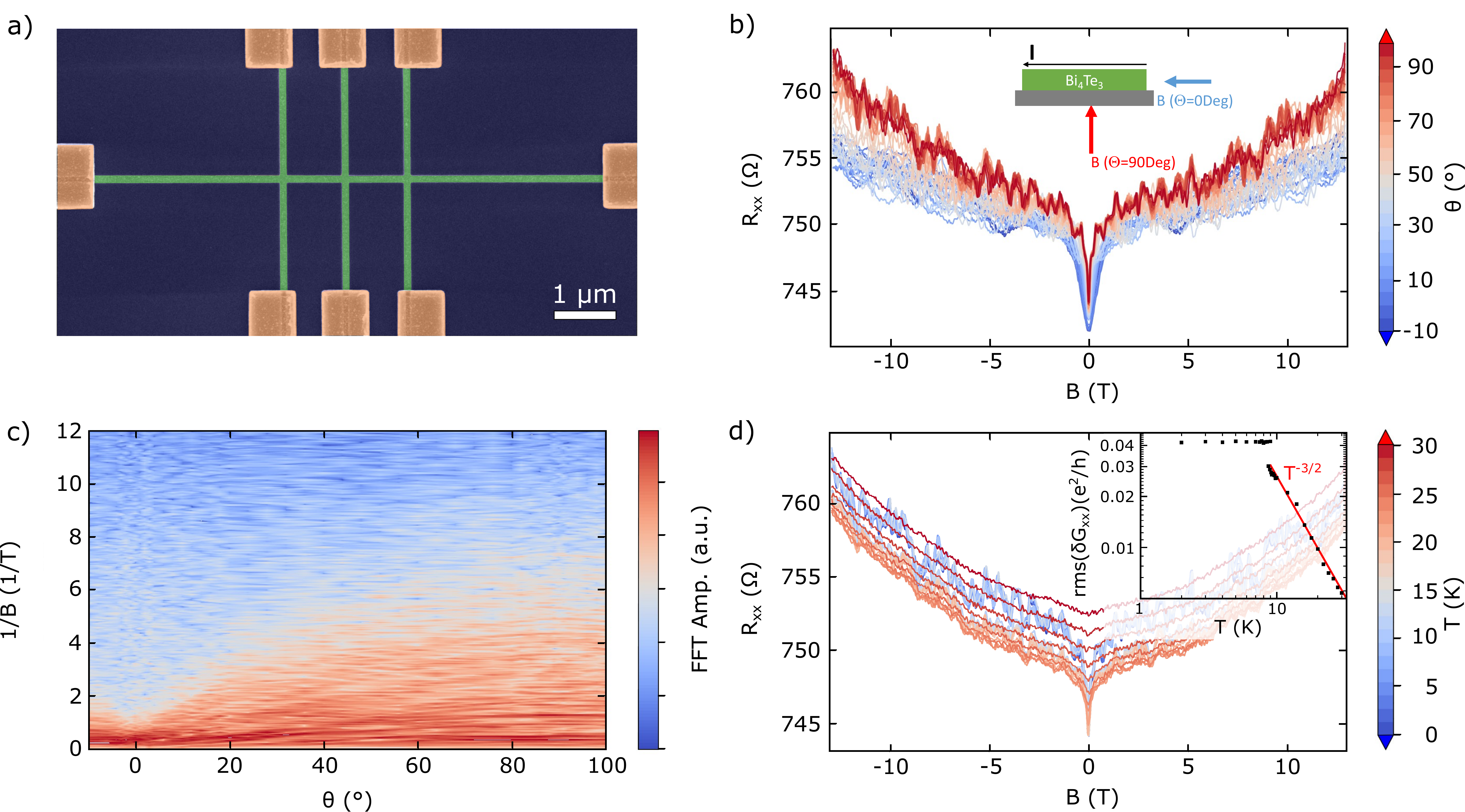}
	\caption{\textbf{Magnetotransport data on Bi$_4$Te$_3$ Hall bars.} a) Layout of the selectively grown Bi$_4$Te$_3$ nano Hall bar investigated. b) Longitudinal magnetoresistance as a function of magnetic field for various tilt angles ($R_{xx}(B, \theta$) of the devices main channel w.r.t the magnetic field. The orientation of the sample is schematically depicted. c) Fast-Fourier-transformation amplitude of the magnetoresistance traces from b) showing high frequent universal conductance fluctuations at a large range of angles in between $15^{\circ} \leq \theta \leq 90^{\circ} $ and low frequent oscillations from coherent states within the nanoribbon cross section for a magnetic field applied parallel to the main axis of the nanoribbon. d) Temperature dependency of the longitudinal magnetoresistance ($R_{xx}(B, T$)) for temperatures in between $1.5\,$K$ \leq T \leq 30\,$K. The inset shows the temperature dependency of the root mean square of the conductance fluctuation amplitude $rms(\delta G_{xx}(T)$}
	\label{figS1}
\end{figure*}

Besides Josephson junctions with a Bi$_4$Te$_3$ weak-link we have fabricated nano Hall bars. Therefore we have selectively deposited Bi$_4$Te$_3$ in nanotrenches that have been arranged in a Hall bar layout with one main nanoribbon and three ribbons symmetrically on each side. The fabricated device is shown as a false color scanning electron micrograph in Fig.~\ref{figS1} a). The nanoribbons have a width of $w=100\,$nm and the spacing in between two side nanoribbons is $L=1000\,$nm.\\
Nano Hall bar devices have been cooled down to $T=1.5\,$K in a variable temperature insert cryostate equipped with a superconducting magnet that can apply magnetic fields up to $B_{\text{max}}=13\,$T. The sample holder insert is equipped with an electromechanical stepper motor. The relative orientation of the magnetic field to the nano Hall bars can be changed from an alignment of the magnetic field parallel to the main nanoribbon axis to a magnetic field oriented perpendicular out-of-plane. From Hall measurements in an out-of-plane applied magnetic field the Hall slope $A_H=dR_{xy}/dB=1.58\,\Omega/$T and subsequently the two-dimensional sheet carrier density $n_{\text{2D}}=(A_H e)^{-1}=3.9\times 10^{14}\,$cm$^{-2}$ have been determined. The Bi$_4$Te$_3$ has a strong metallic character with high charge carrier density and low mobilities $\mu=L \cdot (W R_{xx} n_{\text{2D}} e)^{-1}= 215\,$cm$^2 (V\cdot s)^{-1}$\\
Fig.~\ref{figS1} b) shows the longitudinal resistance of the nano Hall bar for different relative angles of the magnetic field applied to the surface of the substrate. Next to the weak antilocalization feature, typical for 3D bulk as well as 2D surfaces with strong spin-orbit coupling, the magnetoresistance does not change by more then 2.5$\%$ over the whole range of applied magnetic fields $B\leq |\pm 13T|$. In a perpendicular applied magnetic field ($\Theta = 90^{\circ}$, red curve) the magnetoresistance does show a spectrum of universal conductance fluctuations. In Fig.~\ref{figS1} c) the amplitude of individual oscillations frequencies from a fast fourier transformation performed on the data from b), shows a set of prominent frequencies, limited by the phase coherence length ($l_{\phi}=(\phi_0 \cdot f_{\text{max}}) \approx 25\,$nm).\\
The temperature dependent magnetoresistance data in a perpendicular applied magnetic field for temperatures in between $2\,$K $\leq T \leq 30\,$K is shown in Fig.~\ref{figS1} d). For each trace the root mean square of the oscillation amplitude is computed $rms(\delta G_{xx})$ and the values are shown in the inset as a function of temperature. The value for $rms(\delta G_{xx})$ is constant up to a temperature of 9\,K. For higher temperatures the values follow a $T^{-3/2}$ dependency. 

\subsection{Shapiro response measurements}
\begin{figure*}[!ht]
	\centering
\includegraphics[width=\textwidth]{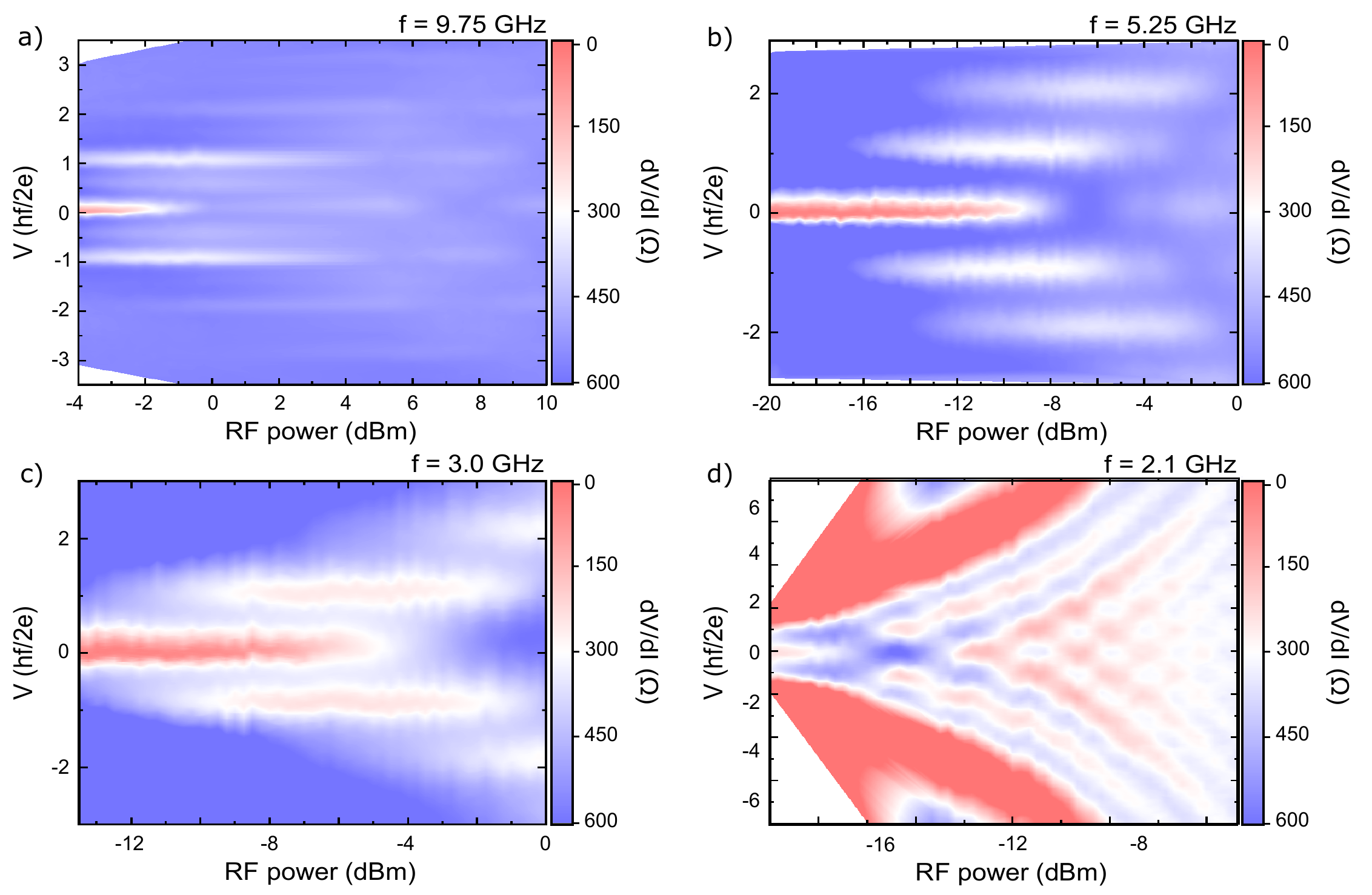}
	\caption{\textbf{Shapiro response of Josephson junction \#3 at different radio-frequencies applied.} a), b), c) and d) show the differential resistance as a function of the radio-frequency excitation amplitude/radio-frequency power ($P$ in dBm) and the potential bias ($V$ in $hf/2e$) of the junction ($dV/dI (P,V)$) at $f=9.75\,$GHz, $f=5.25\,$GHz, $f=3.0\,$GHz and $f=2.1\,$GHz, respectively. The differential resistance is displayed in between values of $dV/dI=0$ (red) and $dV/dI=600\,\Omega$.  Next to Shapiro steps at full integer values of $V=n\cdot hf/2e$, there are sub-integer steps visible for an applied radio-frequency of $f=9.75\,$GHz. }
	\label{figS2}
\end{figure*}
Next to the 500\,nm wide Bi$_4$Te$_3$ Josephson junction characterized in the main text, we have additionally measured a wide junction ($w=1000$\,nm, Junction $\#1$) and a narrow junction ($w=100$\,nm, Junction $\#3$). All the junction parameters are given within the table in the main manuscript. Next to these standard junction characeristics we here show Shapiro step measurements of the narrow junction $\#3$, shown in Fig.~\ref{figS2}. Within a similar range of radiofrequencies applied, as for junction $\#2$ in the main text, we observe a similar behavior w.r.t. the Shapiro step evolution in the differential resistance as a function of the applied RF power applied to and the d.c. potential bias applied across the junction ($dV/dI (P,V)$). For the largest frequency applied of $f=9.75\,$GHz, not only full integer Shapiro steps, but also half-integer Shapiro steps can be observed. This demonstrates that the existence of high coherent ballistic channels do not seem to change with the width of the nanoribbon, as they would in topological insulator nanoribbons, where a quantization of transverse momentum states would alter the surface state dispersion.

\end{document}